\documentclass[12pt]{IEEEtran}
\usepackage{graphicx,color,psfrag}
\usepackage{amsmath}
\usepackage{psfrag}
\usepackage{subfigure}
\usepackage{color}
\usepackage{url}
\usepackage{epsfig}
\usepackage{fancyhdr}

\usepackage[backend=bibtex,style=ieee]{biblatex}
\bibliography{Propagation,WSN,Cellular,Books}
\AtEveryBibitem{\clearfield{url}}
\AtEveryBibitem{\clearfield{doi}}

\newcommand{\Mx}[1]{{\textbf{#1} }}
\newcommand{\Deg}{^{\circ}}
\newcommand{\bMiniPage}[1]{\begin{minipage}{#1}}
\newcommand{\eMiniPage}{\end{minipage}}
\newcommand{\NoI}{\noindent}
\newcommand{\MBx}{\mbox}
\newcommand{\Exp}[1]{E\left\{ #1 \right\}}

\usepackage{ulem}
\begin{document}

\title{Dense Urban Channel Measurements for Utility Pole Fixed Wireless Links}

\author{Michael W.~Wasson, Geoffrey G.~Messier and Devin P.~Smith}

\maketitle

\thispagestyle{empty}

\begin{abstract}
This radio channel measurement campaign characterizes the propagation conditions experienced in a dense urban environment over fixed backhaul links between wireless devices that are mounted on utility or traffic light poles.  The measurements characterize the $2 \times 1$ multiple input single output channel in the 2.45~GHz band for both spatially separated omni antennas and cross polarized directional antennas.  Results presented include both small and large scale channel statistics, antenna correlation coefficient values and the off-broadside rejection achieved with the directional antennas.
\end{abstract}


\IEEEpeerreviewmaketitle

\section{Introduction}

The trend in wireless communications towards moving infrastructure access points off of tall towers and bringing them into the user environment is well established.  Small pico-cell and femto-cell heterogeneous cellular deployment in outdoor urban environments is one of the most popular methods for increasing capacity in next generation cellular networks \cite{ghosh-a-2012}.  In several vehicular ad-hoc network proposals, communication between vehicles and roadside wireless access points is a requirement \cite{hartenstein-h-2008}.  Finally, most sensor network applications involve deploying sensors close to the ground \cite{romer_k1}.

When high density wireless networks are deployed, providing a backhaul network to carry access point traffic is a challenge in most urban centers.  When placing access points on utility poles, street lights or traffic lights, there typically is no wired communications infrastructure to carry backhaul traffic.  In these cases, a backhaul consisting of fixed peer-to-peer wireless links between access points is clearly attractive.

From a propagation perspective, the transmit and receive antennas for these backhaul links will be mounted well below traditional micro-cell antenna height and the transmitter and receiver will both be surrounded by similar scattering environments.  The antennas will be close to vehicular and pedestrian traffic which will influence small scale fading \cite{ahumada-l-2006}.  Most heterogeneous cellular and vehicular network applications will require an access point approximately every city block, which means link distances of approximately 150~m to 200~m.  If considering a North American context, the regular grid layout of most urban centers will allow line of sight (LOS) or near-LOS backhaul channels between access points.  

The outdoor fixed wireless link has received considerable research attention and there have been a number of fixed wireless link propagation measurement campaigns for carrier frequencies close to 2~GHz \cite{greenstein-lj-2009, liou-a-2009, michelson-dg-2009}, 2.5~GHz \cite{erceg-v-2004, gans-mj-2002},  3.5~GHz \cite{ahumada-l-2005, hong-cl-2003a, hong-cl-2003b} and 5~GHz \cite{durgin-gd-1998, domazetovic-a-2003, skentos-nd-2006}. These studies all characterize the fixed wireless access channel where a fixed user terminal is communicating with a base station tower positioned above the scattering environment experienced by the user.  The studies that do adjust antenna height \cite{greenstein-lj-2009, durgin-gd-1998, ahumada-l-2005, gans-mj-2002} adjust the receive antenna to observe the effect of raising the receiver out of the clutter, rather than lowering the transmit antenna into it.  These studies also consider typical macro-cellular link distances ranging from several hundreds of meters to several kilometers.

Many of the existing studies that consider transmit and receive antennas that are both down in the same scattering environment focus on large scale channel effects only.  The below rooftop to below rooftop (BRT to BRT) LOS transmission in \cite{802.16-channel-2007} provides path loss values that would be representative of the fixed wireless backhaul scenario.  Other studies that measure large scale fading values when both transmit and receive antennas are below rooftop include \cite{ahumada-l-2013, xia-h-1994, xia-h-1993, feuerstein-mj-1994} but these campaigns did not set transmit and receive antennas at the same height, as would be typical for a backhaul scenario.  The work in \cite{durgin-gd-2003} is the only study that utilized the same antenna height for the transmitter and receiver.  However, the antenna heights used in \cite{durgin-gd-2003} are only 1.5~m.  The WINNER channel model fixed street level feeder scenario, case B5B in \cite{winner-2005}, is the most relevant to the measurements in this paper and will be used as a point of comparison for both small and large scale fading results.

The contribution of this paper is to present the results of a propagation measurement campaign that characterizes both large and small scale channel effects for fixed wireless links meant to carry backhaul traffic for heterogeneous cellular, vehicular and sensor network infrastructure in a dense urban environment.  Both the transmit and receive antennas are mounted at the same height as would be typical for a backhaul scenario.  The results from this campaign will complement existing fixed wireless propagation work that has focused on access scenarios with cellular style towers and very long link distances.  Dual antenna multiple input/single output (MISO)  measurements are collected for both spatially separated omni antennas and dual-polarized directional antennas.  The results will demonstrate how antenna pattern affects channel statistics and the correlation between antennas for the urban wireless backhaul scenario.

An additional contribution of this paper is to focus specifically on the properties of the propagation environment that will affect the ability of a wireless backhaul unit to simultaneously support several spatial data streams.  In a grid-style peer-to-peer backhaul network where radio nodes located at each intersection are serving as relays for other nodes, it will be common for one node to be communicating with devices in  multiple directions.  One way to handle these streams simultaneously is to use multiple antenna techniques: either spatial multiplexing or directional antenna space division multiple access (SDMA).  To help evaluate the feasibility of SDMA, this study will measure the spatial rejection achieved in the off-broadside direction using directional antennas.  The feasibility of spatial multiplexing will be determined by evaluating the correlation between spatially separated omni antennas both in the broadside and off-broadside directions.

Section~\ref{sec:meas} discusses the radio channel measurement equipment, measurement collection procedure and the environment where the measurements were captured.  The method used to analyze the propagation measurements is discussed in Section~\ref{sec:analysis} and the results of this analysis are presented in Section~\ref{sec:results}. Concluding remarks are made in Section~\ref{sec:concl}.

\section{Propagation Measurements}
\label{sec:meas}

In this section, Subsection~\ref{ssec:equip} describes the radio channel measurement equipment and Subsection~\ref{ssec:proc} discusses how the equipment is used in the propagation environment to capture wireless channel response data.

\subsection{Equipment}
\label{ssec:equip}

The measurement apparatus used for this study collects $2 \times 1$ MISO channel impulse response measurements.  The transmit unit consists of two field programmable gate array (FPGA) software radio boards installed inside a commercial pico-cell base station enclosure. Each radio is connected to a different transmit antenna port and both are locked to a common time and carrier frequency reference.  Each radio generates the same 25~Mchip/sec, length $2^{19}-1$ maximal length pseudo-random noise (PN) sequence \cite{proakis_jg1} and utilizes an offset of $2^{18}$ chips to ensure orthogonality. The offset sequences are denoted $p_1[n]$ and $p_2[n]$.  After digital to analog conversion, the signals have an approximate 3~dB double sided bandwidth of 25~MHz.  They are amplified to a maximum transmit power of 36~dBm equivalent isotropically radiated power (EIRP) and transmitted at a carrier frequency of 2.4724~GHz.

The receive unit contains a single FPGA software radio board that implements two real time correlators, one correlating the received signal with $p_1[n]$ and the other with $p_2[n]$.  The correlation is performed over 100,000 chips so that a $2 \times 1$ MISO channel impulse response is captured every 4~ms for a measurement capture rate of 250~Hz.  The channel response from antenna $i$ is represented by matrix $\Mx{h}_i \in {\cal C}^{N \times L}$ where $N$ is the number of channel impulse responses captured and $L$ is the number of discrete channel taps.  The element at row $n$ and column $l$ of $\Mx{h}_i$ is denoted $h_i(n,l)$.

Time and frequency synchronization for the transmit and receive units is provided by SIM940 rubidium clock references \cite{sim940}.  The references provide both 1 pulse per second and 10~MHz reference signals.  They are synchronized to each other after a warm up period and are battery powered to ensure synchronization is maintained during the measurements.

Measurements are collected using either omni-directional antennas or directional antennas. For the omni antenna measurements, broadside is defined as perpendicular to the line through the two transmit antenna elements.  Broadside for the directional antenna measurements is perpendicular to the panel antenna face in the direction of maximum antenna gain.  Off broadside for both cases is 90$\Deg$ off of the broadside direction in azimuth.

The omni antennas are the W1030 quarter wavelength dipole antennas manufactured by Pulse Electronics \cite{w1030-antenna-2014}.  The antenna has a gain of 2.0~dBi and an omni-directional horizontal radiation pattern.  The separation between the two transmit omni dipole antennas is 23~cm.  This separation is dictated by the location of the antenna ports on the pico-base station enclosure used to house the transmitter software radios.

The directional antennas are the MT-344048/ND cross polarized panel antennas manufactured by MTI Wireless \cite{mt344048-antenna-2014}.  The panel antennas mount on the front of the base station enclosure.  They have a 30$\Deg$ beamwidth and have antenna gain patterns that satisfy the ETSI EN 302 326-3 cross polarized directional antenna standard. This is an antenna standard created by the European Telecommunications Standards Institute governing antenna requirements for fixed wireless applications.  The details on this antenna standard, including quantitative values for the antenna gain masks can be found in \cite{etsi-ants}.   

It is important to verify that the system has sufficient signal to noise ratio to ensure measurement noise will not be mistaken as fading when measuring very stable wireless links \cite{ahumada-l-2005}.  To this end, the measurement transmitter and receiver were connected with a cable that included 80~dB of attenuation.  The resulting signal to noise ratio of the system, as defined in \cite{ahumada-l-2005}, was 44.3~dB.  The methods in \cite{ahumada-l-2005} can be used to show this is high enough for measurement noise to cause only negligible error for the channels measured in this study.

The parameters for the measurement equipment section are summarized in Table~\ref{tb.equip}.

\begin{table}[htbp]
  \centering
  \begin{tabular}{|c|c|}
    \hline
    Parameter & Value \\
    \hline\hline
    Carrier Frequency & 2.4724~GHz \\
    Signal Bandwidth & 25~MHz \\
    Maximum Transmit Power & 36~dBm~EIRP \\
    MISO Configuration & 2$\times$1 \\
    Impulse Response Capture Rate & 250~Hz \\
    Omni Antenna Gain & 2~dBi \\
    Directional Antenna Gain & See \cite{etsi-ants}\\
    \hline
  \end{tabular}
  \caption{Measurement equipment parameters.}
  \label{tb.equip}
\end{table}

\subsection{Measurement Procedure}
\label{ssec:proc}

\begin{figure}[htbp]
\centering
\bMiniPage{1.5in}
 \centering
 a) Transmitter\\[2mm]
 \includegraphics[width=1.4in]{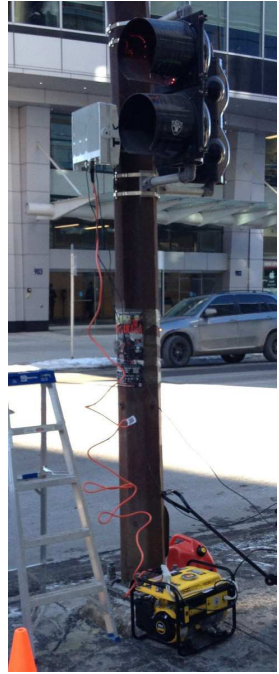}
\eMiniPage%
\bMiniPage{1.5in}
 \centering
 b) Receiver\\[2mm]
 \includegraphics[width=1.4in]{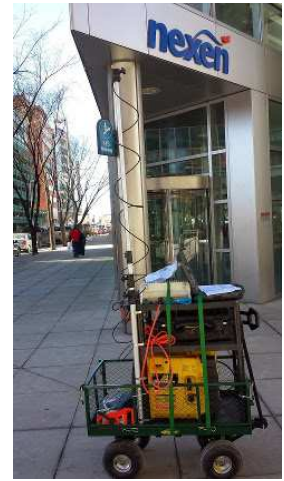}
\eMiniPage
\caption{Measurement equipment.}
\label{fg.setup}
\end{figure}

At the start of each measurement session, the transmit unit is attached to a traffic light pole, as shown in Fig.~\ref{fg.setup}a. The target mounting height for the transmit unit is 3~m but this would vary $\pm$0.25m from site to site depending on the configuration of the pedestrian walk lights on the traffic light pole.  The receive equipment, shown in Fig.~\ref{fg.setup}b, is placed in a wheeled cart equipped with an antenna mast.  In order for the measurements to be representative of a peer-to-peer wireless backhaul scenario, the mast is adjusted so that the receive antenna is the same height as the antennas on the transmit unit.

\begin{figure}[htbp]
\centerline{\includegraphics[width=3.5in]{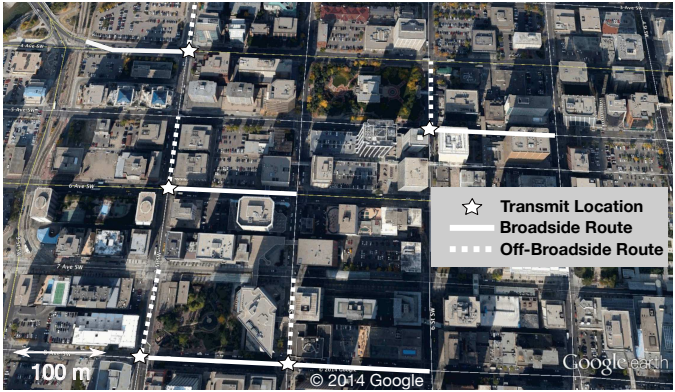}}
\caption{Measurement locations.}
\label{fg.map}
\end{figure}

A map of the measurement environment is shown in Fig.~\ref{fg.map}.  The area is dominated by office buildings that range in height from 20 to 50 storeys.  There is either single or double lane vehicular traffic on each street with typical vehicular speeds not exceeding 50~km/hour.  The streets also include broad pedestrian sidewalks.  All measurements were collected during the business day so that vehicular and pedestrian traffic levels ranged from moderate to heavy.

In Fig.~\ref{fg.map}, the stars indicate the different locations where the transmit unit was mounted to a traffic light pole.  For each of these locations, the receiver cart is used to collect data while moving down the street in both the broadside and off-broadside directions, as indicated in Fig.~\ref{fg.map}.  A single trip down the street with the receiver cart is referred to as a measurement {\em route}.  All routes start with the cart no more than 5~m away from the transmit unit and end with the cart one block away from the transmit unit, as indicated by the solid and dotted lines in Fig.~\ref{fg.map}.

For each transmitter location, the broadside and off-broadside directions are measured once for the omni antenna case and once for the directional antenna case using the following procedures.

\subsubsection{Omni Antenna Measurement Procedure}
\label{ssec:omni-proc}

When moving the receive cart along a particular route,  a pair of $2\times 1$ measurement channel impulse response matrices, $\Mx{h}_1$ and $\Mx{h}_2$, are captured at measurement locations spaced by approximately 5~m.  This spacing would vary somewhat due to obstructions on the sidewalk. Approximately 20 of these measurements are captured for the off-broadside routes and 35 measurements for the broadside routes. 

When the cart arrives at measurement location, the receive correlator is activated and the apparatus begins to capture MISO channel responses at a rate of 250~Hz, as described in Section~\ref{ssec:equip}.  In order to capture small scale fading variation due to spatial offsets, the receiver cart is moved within a square of at least 0.5~m $\times$ 0.5~m while the MISO channel impulse responses are being captured.  The size of this 0.5~m $\times$ 0.5~m measurement grid falls within the range of grid sizes reported for capturing small scale fading in a local neighbourhood \cite{chiu-s-2010,chee-kl-2012}.  The measurements captured using this approach are referred to as {\em spatial measurements}.

To characterize small scale fading variation due to moving objects in the environment, measurements were also collected while the transmit and receive antennas were completely stationary.  These are referred to as {\em temporal measurements}.  One temporal measurement is captured at the end of each measurement route where the separation between transmitter and receiver was at its maximum.  The duration of each of these measurements was 5-6~minutes and the measurement capture rate was 250~Hz.  Any channel variation longer than this 5-6~minute duration will not be reflected in the analysis presented in this paper.

\subsubsection{Directional Antenna Measurement Procedure}
\label{ssec:dir-proc}

In order to reflect a typical backhaul scenario, it is important that the transmit and receive directional antennas are pointed directly at each other for all propagation measurements.  This alignment is difficult to maintain while moving the cart from side to side in a square grid.  The risk is that the temporary misalignment of the beams due to cart motion may cause artificial fading.

To address this concern, the receive correlator is enabled at the start of a measurement route and MISO channel impulse responses are captured continuously as the cart is moved in a straight line to the end of the route.  This allows the antenna orientation to be properly maintained.  Spatial small scale fading variations are characterized by dividing the measurement track into 2~m segments and processing all channel impulse responses captured within a segment to determine spatial small scale fading statistics in that local area. 

Unlike the omni antenna, the directional panel antenna used on the receive cart has two spatial polarizations.  As a result, a separate set of measurements are collected along each route for each of the polarizations.

\section{Measurement Analysis}
\label{sec:analysis}

In this section, the procedure for extracting large scale channel information from the measurements is described in Section~\ref{ssec:lscale} and the small scale fading analysis is described in Section~\ref{ssec:sscale}.

\subsection{Large Scale Channel Effects}
\label{ssec:lscale}

The large scale channel effect analysis consists of estimating both the path loss exponent and shadowing distribution for each measurement route.  The first step is to determine the average received power from transmit antenna $i$ at a particular measurement location according to

\begin{equation}
P_i = \frac{1}{N}\sum_n\sum_l |h_i(n,l)|^2
\label{eq.AvgRxPwr}
\end{equation}

\NoI
For the omni antenna case, values of $P_i$ are calculated by performing the averaging in (\ref{eq.AvgRxPwr}) on each $0.5 \times 0.5$~m spatial measurement described in Section~\ref{ssec:omni-proc}.  For the directional antennas, a value of $P_i$ is calculated for each of the 1~m measurement track segments described in Section~\ref{ssec:dir-proc} by applying (\ref{eq.AvgRxPwr}) to all the channel impulse responses captured in each segment.

As described in \cite{rappaport_ts1}, the path loss exponent is calculated as the slope of the best fit line on a log-log plot of average received power values versus distance between the transmit and receive radios.  The shadowing values are the deviation of each average received power point to that best fit line.  Separate path loss exponents and shadowing average power deviation values are calculated for each transmit antenna and for the two different polarizations of the directional receive antenna.  

For this study, a separate line of best fit is calculated for each measurement route.  This produces several path loss exponent values and also results in a tighter shadowing distribution than if a single line of best fit was calculated for all received power values from all streets.  This is desirable since this tighter distribution more accurately reflects shadowing conditions on a particular street.  Note that a single shadowing distribution is still calculated by aggregating average power deviation values from all streets into a single histogram.  However, these deviation values are determined using the best fit lines calculated on a per street basis.

\subsection{Small Scale Channel Effects}
\label{ssec:sscale}

To analyze spatial small scale fading variation, all channel impulse responses captured in a $0.5 \times 0.5$~m grid, for the omni case, or a 1~m linear segment, for the directional case, are normalized to have unit average power.  To analyze temporal small scale variation, the temporal measurement is divided into 60~s segments and each segment is normalized and analyzed separately.  The responses are then transformed to the frequency domain by taking the discrete Fourier transform of each impulse response, $h_i(n,l)$, along index $l$.  Let $H_i(n,w)$ denote the transformed impulse response where $w$ is the discrete frequency index and the definitions of $n$ and $i$ are unchanged.  The channel frequency response matrix is defined as $\Mx{H}_i \in {\cal C}^{N \times L}$, where the $n$th row is the Fourier transform of the channel impulse response captured at time $n$.

If $W_{\rm coh}$ is the discrete coherence bandwidth of the channel and $W_{\rm sig}$ is the discrete bandwidth of the measurement signal, then $R = \lfloor W_{\rm sig}/W_{\rm coh} \rfloor$ is the number of channel frequency response values with independent small scale fading.  Since $N$ fading values are captured for each of these frequency points, this allows the creation of  a $1 \times NR$ {\em fading vector} for the $i$th transmit antenna defined as $\Mx{e}_i = [\ |H_i(0,0)|\ \ldots\ |H_i(N-1,0)|\ |H_i(0,W_{\rm coh})| \ldots\ |H_i(N-1,RW_{\rm coh})|\ ]$.

Antenna correlation is calculated by determining the correlation coefficient between $\Mx{e}_1$ and $\Mx{e}_2$.  While this is only a $2 \times 1$ system, it is important to point out that the correlation between the two antenna elements at the transmitter would be representative of the correlation between elements of a larger array with the same element spacing.

To quantify the severity of fading in the channel, the amount of fading (AF) metric is used \cite{alouini-2005}.  AF is defined as $AF = \MBx{variance}\left(\Mx{e}_i^2\right)/\Exp{\Mx{e}_i^2}^2$. As will be discussed in Section~\ref{sec:results}, there are several instances where the fading of $\Mx{e}_i$ cannot be modeled as Ricean.  AF is therefore a more appropriate fading metric since it is independent of a particular fading distribution.

Two different types of AF values are determined.  {\em Temporal AF} values are calculated from measurements where both the transmit and receive antennas are stationary and channel variation is due to movement of scattering objects in the environment.  {\em Spatial AF} values are calculated from measurements where the receive antenna is moved about a small local area of a few wavelengths during the measurement.  In the spatial case, fading is also caused by the motion of the antenna and scatterers.

It is important for fixed wireless network designers to understand both the spatial and temporal fading.  The temporal variations are clearly the fluctuations that the wireless link must contend with after deployment.  However, spatial fading characterizes the possible variation in received power that may occur due to small offsets in the mounting position for the antenna.  While spatial variations are constant once the antenna is mounted, the network designer must still allow a fading margin in the link budget calculation to account for them.  Basing a link budget fading margin on temporal fading alone would result in transmit powers that are too low.

\section{Measurement Results}
\label{sec:results}

In this section, Subsections~\ref{ssec:lsanalysis} and \ref{ssec:ssanalysis} present the large scale and small scale analysis results, respectively.  Subsection~\ref{ssec:rejanalysis} discusses the measurements used to evaluate the feasibility of using multiple antennas to support simultaneous communication in the broadside and off-broadside directions.

\subsection{Large Scale Path Loss Analysis}
\label{ssec:lsanalysis}

\begin{figure}[htbp]
\centerline{\includegraphics[width=3.5in]{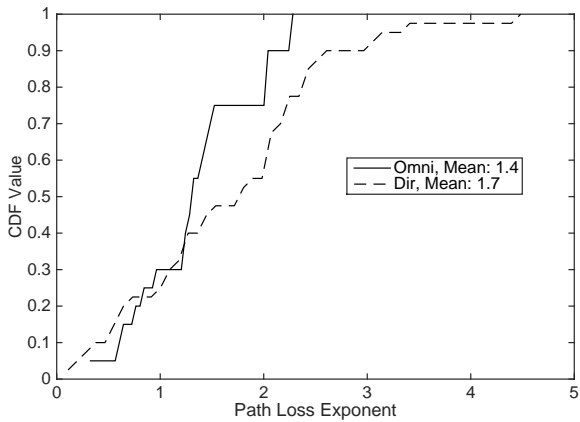}}
\caption{Path loss exponent values.}
\label{fg.pathloss}
\end{figure}

As discussed in Section~\ref{ssec:lscale}, separate path loss exponents are determined for all intersections.  Cumulative distribution function (CDF) plots of the path loss exponent values for both the omni and directional antennas in the broadside direction are shown in Fig.~\ref{fg.pathloss}.

We can note from Fig.~\ref{fg.pathloss} that the path loss exponent values are quite low with the means of both antenna types close to but below 2.  This is consistent with path loss values in \cite{xia-h-1994, xia-h-1993, feuerstein-mj-1994, winner-2005} and is due to the urban canyon acting as a waveguide that serves to reduce path loss exponent.   A slightly higher pathloss exponent was observed in \cite{ahumada-l-2013} but in that study, the authors note that even their line of sight measurements were typically obscured by trees.  We also note that the directional antennas exhibit a slightly higher path loss exponent since omni signals are more diffuse in the environment and are less affected by obstructions in the direct line of sight between transmitter and receiver.

\begin{figure}[htbp]
\centerline{\includegraphics[width=3.5in]{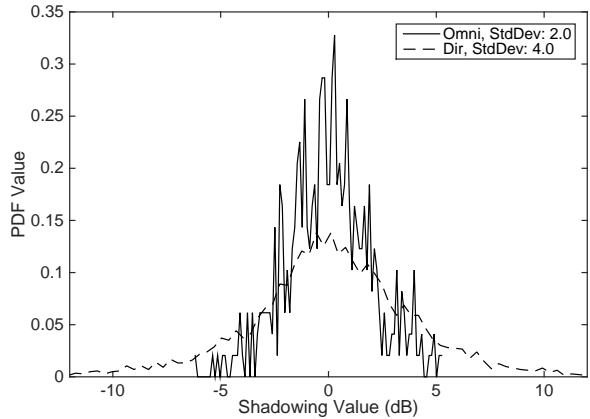}}
\caption{Shadowing distributions.}
\label{fg.shadow}
\end{figure}

Fig.~\ref{fg.shadow} shows the shadowing distribution for the two antenna types.  Both antenna types experience shadowing that is approximately log-normal with the directional antenna experiencing a larger standard deviation.   The smaller standard deviation for the omni antenna is again due to the signal being more diffuse in the propagation environment.  The shadowing standard deviation for both cases is comparable to the B5B scenario in \cite{winner-2005}.

\begin{figure}[htbp]
  \centerline{\includegraphics[width=3.5in]{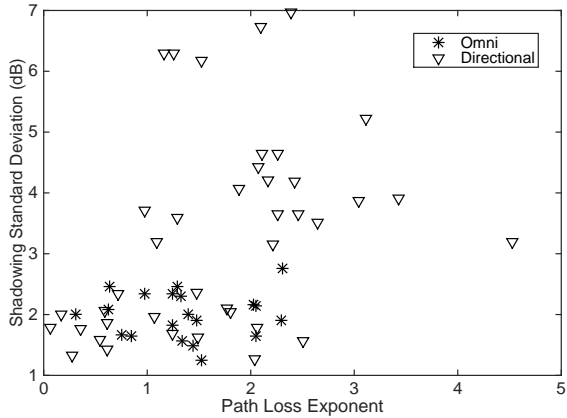}}
  \caption{Path loss exponent versus shadowing standard deviation.}
  \label{fg.nAndShdwCorr}
\end{figure}

As described in Section~\ref{ssec:lscale}, the path loss and shadowing values are calculated on a per-measurement route basis.  Since some streets with more obstructions may be expected to have both a higher path loss exponent and a higher shadowing variation, it is possible that some correlation may exist between the two quantities.  This can be determined from the scatterplot in Fig.~\ref{fg.nAndShdwCorr} which shows the path loss exponent for each measurement route plotted versus the shadowing standard deviation for that route.

Fig.~\ref{fg.nAndShdwCorr} shows little relationship between path loss and shadowing standard deviation for the omni antenna but a stronger correlation for the directional antenna.  To support this, a calculation of correlation coefficient between the two values shows that path loss exponent and shadowing standard deviation have a correlation coefficient of 0.048 for the omni case but 0.39 for the directional case.  The higher correlation for the directional case is most likely due to the sensitivity of narrow beams to obstructions in the line of sight path between transmitter and receiver.

\subsection{Broadside Small Scale Fading Analysis}
\label{ssec:ssanalysis}

As described in Section~\ref{ssec:sscale}, the first step in assembling the fading vectors for analysis is to calculate the coherence bandwidth and determine the number of frequencies in the channel response with independent fading.  The mean coherence bandwidths for the omni and directional antennas, calculated as described in \cite{rappaport_ts1}, are 2.33~MHz and 5.50~MHz, respectively.  The higher coherence bandwidth for the directional antenna is expected since the beam pattern will reject some scatterers and reduce the delay spread of the channel. 

Next, each small scale fading vector, $\Mx{e}_i$, is subjected to a Chi-square goodness-of-fit test \cite{navidi_wc1} to determine if it fits a Ricean distribution.  A significance level of 5\% is used, meaning there is a 5\% chance that measurements that are actually Ricean will be discarded.  The percentage of measurements that fit a Ricean distribution is 100\% and 88\% for the omni antenna spatial and temporal scenarios, respectively.  For the directional antenna, 38\% and 68\% of the spatial and temporal measurements, respectively, satisfy the Ricean fit.  The directional spatial measurements, in particular, were a poor fit to the Ricean distribution primarily due to the straight line measurement procedure described in Section~\ref{ssec:dir-proc}.  The reduced number of spatial samples in each measurement results in distributions concentrated around a small number of discrete fading values.

\begin{figure}[htbp]
\centerline{\includegraphics[width=3.5in]{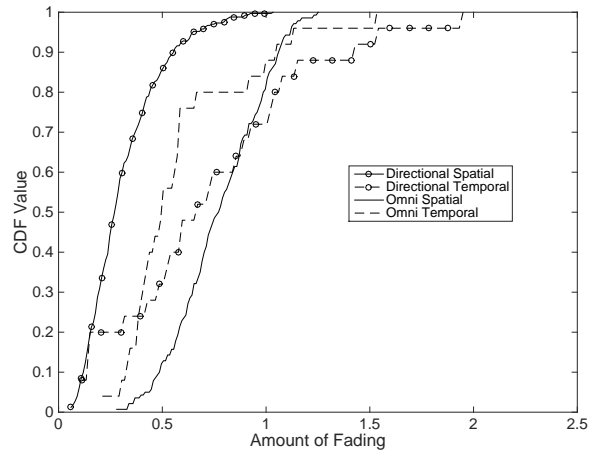}}
\caption{Omni and directional AF distributions.}
\label{fg.AF}
\end{figure}

Fig.~\ref{fg.AF} shows spatial and temporal linear AF CDFs for the omni and directional antennas.  The mean linear AF values for the omni spatial, omni temporal, directional spatial and directional temporal scenarios are 0.778, 0.592, 0.313 and 0.730, respectively.  As expected, the omni spatial channel exhibits more severe fading than the directional spatial channel since the omni antenna gain pattern admits more scattering rays than the directional pattern.  While the directional temporal fading is slightly more severe than the omni temporal fading, fading for the two scenarios is still quite close.  Temporal fading is expected to be similar for the two antenna types since most of the scattering objects are close to the line of sight path due to the large transmitter/receiver separations used for the temporal measurements.  At this distance, the directional antenna pattern is not narrow enough to reject the signals from these scatterers.  Finally, when the B5B K-factor in \cite{winner-2005} is translated to AF, the result is 0.174.  This indicates that the fading measured in this campaign for all scenarios is considerably more severe than that assumed by the WINNER model.

\begin{figure}[htbp]
\centerline{\includegraphics[width=3.5in]{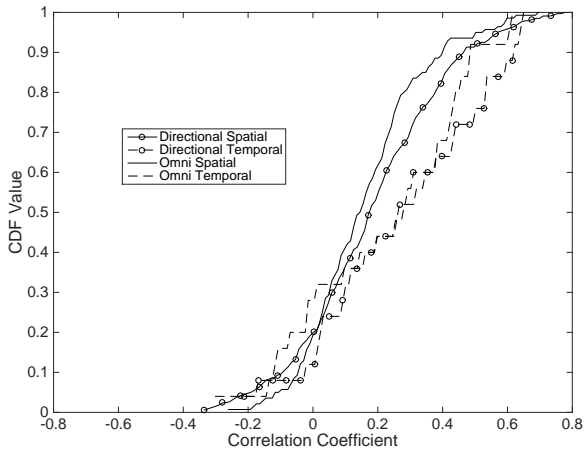}}
\caption{Omni and directional correlation coefficient in the broadside direction.}
\label{fg.brdCorr}
\end{figure}

Fig.~\ref{fg.brdCorr} shows correlation coefficients for all measurement and antenna types in the broadside direction with the directional antenna curves marked with circles.  The results in Fig.~\ref{fg.brdCorr} indicate very similar correlation performance in all cases in the broadside direction which suggests that both spatial and polarization diversity are good options in this environment.  It should be noted that no real dependence with distance was observed in the correlation coefficient values.  This is in contrast with \cite{erceg-v-2004} which did observe some decrease in antenna correlation with distance.  The primary reason for the difference here is that the transmitter/receiver separation distances in this study are much smaller than \cite{erceg-v-2004}.  The environment is also very uniform along each measurement route due to the low antenna heights and urban canyon environment.

\subsection{Broadside and Off-broadside Propagation Conditions}
\label{ssec:rejanalysis}

In this subsection, propagation conditions are compared in the broadside and off-broadside directions.  These results will shed some light on the feasibility of using multiple antenna techniques to allow a node to simultaneously communicate in these two directions.

First, the ability of the directional antenna to reject signals from the off-broadside direction is investigated.  A high degree of rejection would indicate that SDMA using fixed beams is a viable option.  The analysis starts by comparing the difference in average received power levels for the broadside and off-broadside directions.  

Let $P_{xy,B}$ and $P_{xy,OB}$ denote the average received powers in the broadside and off-broadside directions, respectively, measured for the same transmitter/receiver separation, $d$.  The subscripts $x$ and $y$ indicate transmit and receive antenna polarizations, respectively, so that $x,y \in \{ V, H \}$.  Cross-polar rejection in the broadside direction is quantified by the distribution of all $P_{VV,B} - P_{VH,B}$ and $P_{HH,B} - P_{HV,B}$ values calculated for all $d$. Cross-polar rejection in the off-broadside direction is the distribution of all $P_{VV,B} - P_{VH,OB}$ and $P_{HH,B} - P_{HV,OB}$ values for all $d$.  Finally, co-polar rejection in the off-broadside direction is the distribution of all $P_{VV,B} - P_{VV,OB}$ and $P_{HH,B} - P_{HH,OB}$ values for all $d$.

\begin{figure}[htbp]
\centerline{\includegraphics[width=3.5in]{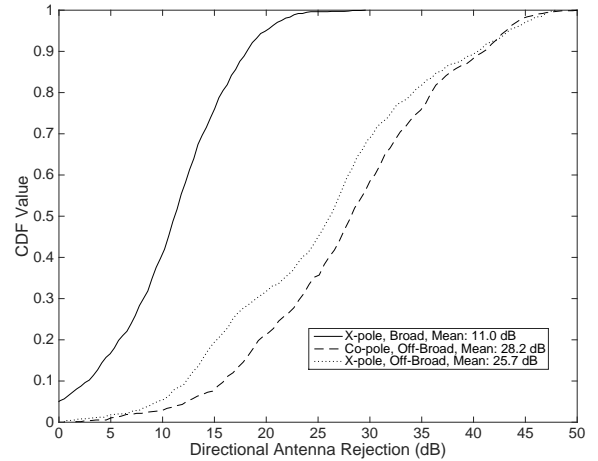}}
\caption{Directional antenna rejection.}
\label{fg.reject}
\end{figure}

A plot of the CDF of the antenna rejection values is shown in Fig.~\ref{fg.reject}.  First, the average cross-polar rejection of 11.0~dB in the broadside direction is close to the 9~dB cross-polar rejection value for the B5B scenario in \cite{winner-2005}.  This verifies the accuracy of the measurements.  The figure also indicates very similar rejection in the off-broadside direction for both the co-polar and cross-polar measurements.  Since almost all measurement locations in the off-broadside direction were obscured, the off-broadside signals experienced heavy scattering that served to spread the signal across both polarizations.  As a result, the primary mode of signal rejection for both cases was the directional antenna pattern.

\begin{figure}[htbp]
\centerline{\includegraphics[width=3.5in]{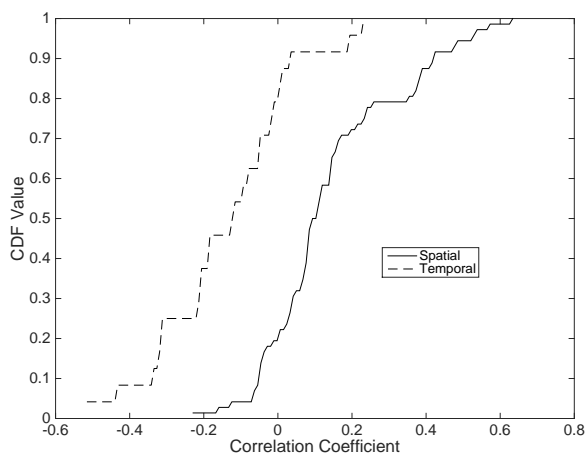}}
\caption{Off-broadside omni fading correlation coefficient.}
\label{fg.oBrdCorr}
\end{figure}

Spatial multiplexing is the second option for simultaneously accommodating broadside and off-broadside communication.  Fig.~\ref{fg.oBrdCorr} shows the spatial and temporal correlation coefficients in the off-broadside direction for the omni antenna measurements.  Omni antenna results are shown based on the assumption that the spatial multiplexing would be implemented using a series of spaced antenna array elements.

The figure shows that both spatial and temporal fading spatial correlation is low enough to support a spatial multiplexing scheme.  It should be emphasized that these low correlations are achieved using linearly spaced antenna elements that are approximately in line with the direction of transmission.  The figure shows that the temporal correlation coefficient is noticeably lower than the spatial correlation.  This is due primarily to the fact that the temporal measurements are collected only at the far end of each off-broadside measurement route where the direct path between the transmitter and receiver is typically obscured.

\section{Conclusion}
\label{sec:concl}

This paper has presented a propagation study for fixed wireless links in a dense urban environment where the antennas are mounted on utility poles or street lights.  The antenna heights are much lower than traditional micro-cellular base stations and the transmit and receive units experience similar physical scattering environments.

Large scale fading analysis indicates that the wireless link is best characterized with a very low path loss exponent and shadowing standard deviation relative to traditional dense urban cellular propagation models.  This reflects both the near-LOS conditions of the links and the per-street method used to calculate the shadowing values.  Antenna type does have an effect on large scale fading with directional antennas experiencing larger shadowing variation.

Small scale fading analysis indicates moderate fading with the majority of measurements experiencing less than Rayleigh fading severity.  Directional antennas successfully reduce fading severity but only in the spatial measurement case.  In the broadside direction, both polarization and spatial diversity would offer equivalent performance.

Finally, multi-antenna techniques show some promise for supporting simultaneous communication in the broadside and off-broadside directions.  A spatial multiplexing approach would be more attractive due to the very low fading correlation observed in the off-broadside direction. SDMA is also attractive since directional antennas demonstrate good rejection in the off-broadside direction with approximately equivalent performance for co-polarized and cross-polarized transmission.

\section*{Acknowledgment}

This work was supported by TEKTELIC Wireless Communications Inc., SRD Innovations Inc. and the Canadian Natural Sciences and Engineering Research Council (NSERC).

\printbibliography

\begin{IEEEbiography}
  [{\includegraphics[width=1in,height=1.25in,clip,keepaspectratio]{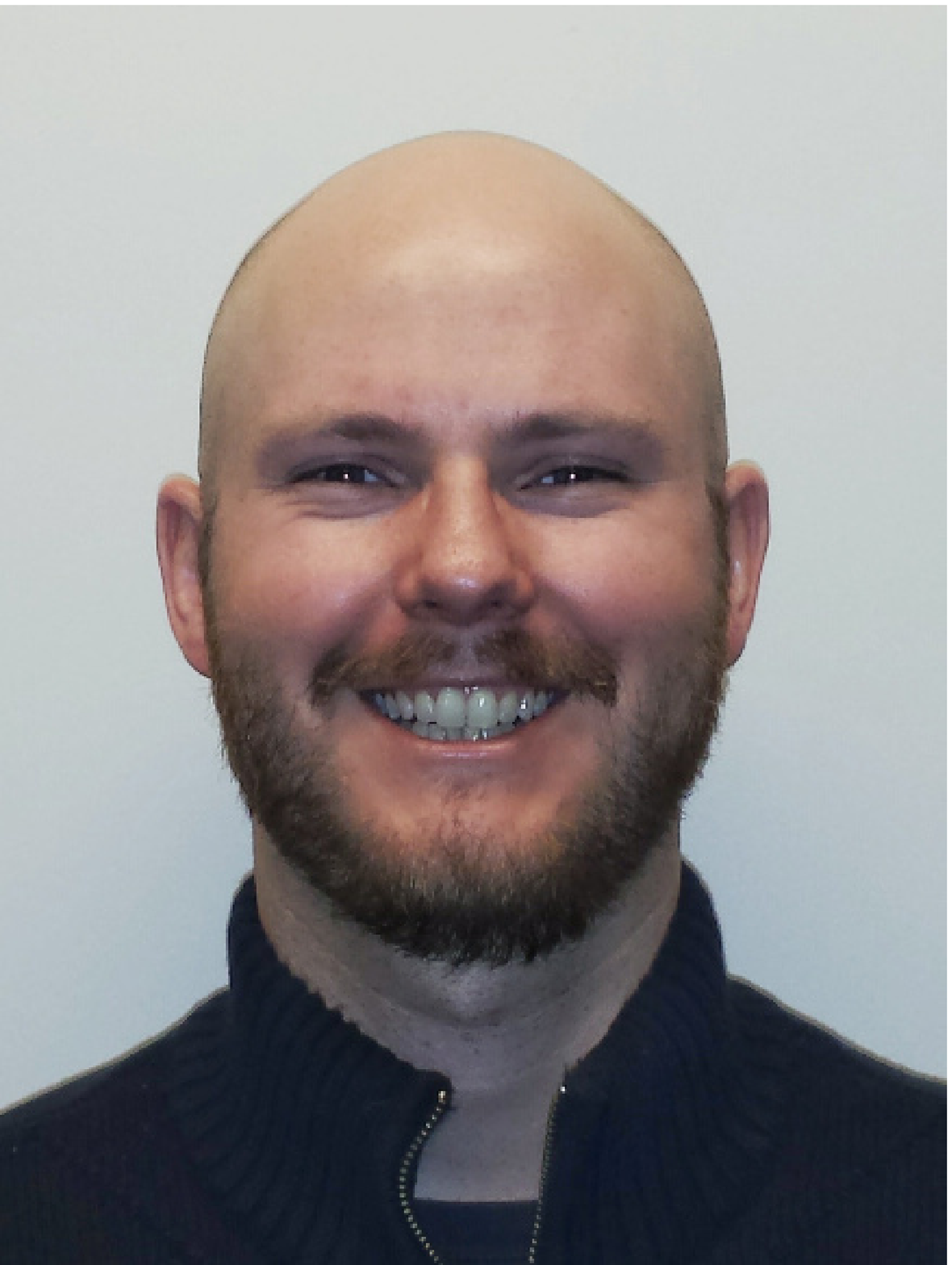}}]{Michael Wasson}
Michael W. Wasson received a B.Sc. degree in electrical engineering in 2010, and is expecting to receive his M.Sc. in electrical engineering in 2016, both from the University of Calgary in Alberta, Canada. 
\end{IEEEbiography}

\begin{IEEEbiography}
  [{\includegraphics[width=1in,height=1.25in,clip,keepaspectratio]{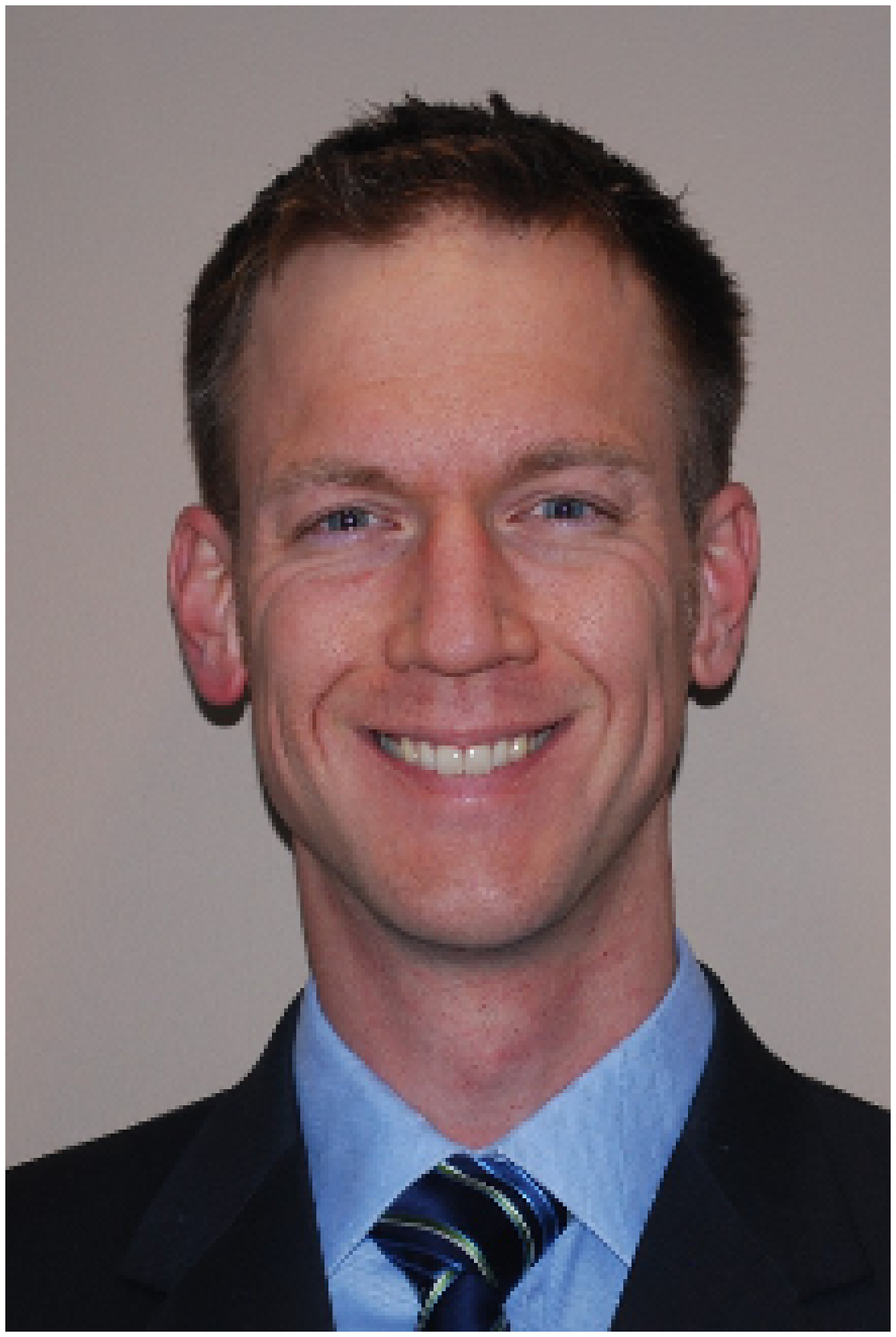}}]{Geoffrey Messier}
Geoffrey Messier (S'91 - M'98) received his B.S. in Electrical
Engineering and B.S. in Computer Science degrees from the University of
Saskatchewan, Canada with great distinction in 1996.  He received his
M.Sc. in Electrical Engineering from the University of Calgary, Canada
in 1998 and his Ph.D. degree in Electrical and Computer Engineering from
the University of Alberta, Canada in 2004.

From 1998 to 2004, he was employed in the Nortel Networks CDMA Base
Station Hardware Systems Design group in Calgary, Canada.  At Nortel
Networks, he was responsible for radio channel propagation measurements
and simulating the physical layer performance of high speed CDMA and
multiple antenna wireless systems.  Currently, Dr. Messier is a
Professor in the University of Calgary Department of
Electrical and Computer Engineering.  His research interests include
data networks, physical layer communications and communications
channel propagation measurements.
\end{IEEEbiography}

\begin{IEEEbiography}
  [{\includegraphics[width=1in,height=1.25in,clip,keepaspectratio]{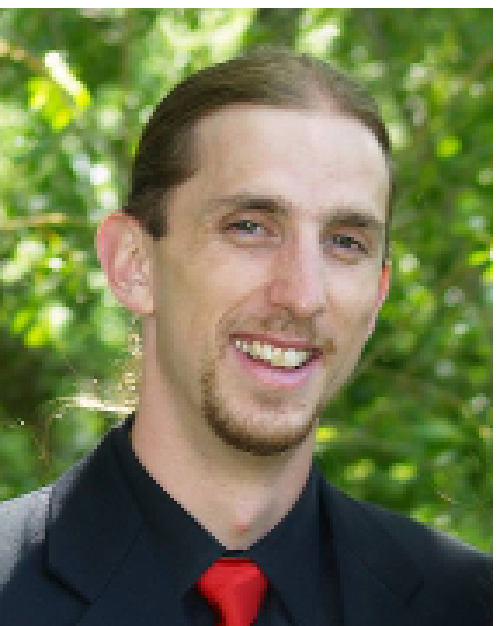}}]{Devin Smith}
Devin Peter Smith was born in Calgary, Alberta, Canada, in 1988. He received his BSc. in Computer Engineering from the University of Calgary in 2012, and is expecting to complete his MSc. in Electrical Engineering in 2015.  Currently, his main research interests have been focused on wireless propagation and sensor networks, specifically within forest environments.
\end{IEEEbiography}

\end{document}